\documentclass[prl,aps,twocolumn,superscriptaddress,nofootinbib]{revtex4}

\usepackage{epsfig}
\usepackage{amsfonts}
\usepackage{latexsym}
\usepackage{amsmath}
\usepackage{amssymb}
\usepackage{slashed}
\usepackage{longtable}

\usepackage{graphicx}
\usepackage{hyperref}
\usepackage{pstool}

\numberwithin{equation}{section}

 \def\p{\partial}

\newcommand{\bea}{\begin{eqnarray}}
\newcommand{\eea}{\end{eqnarray}}
\newcommand{\be}{\begin{equation}}
\newcommand{\ee}{\end{equation}}
\newcommand{\ba}{\begin{align}}
\newcommand{\ea}{\end{align}}

\def\Or[#1]{{\text{O}}\left({#1}\right)}
\def\dotl[#1,#2]{\left\langle #1, #2 \right\rangle}
\def\dotlb[#1,#2]{[ #1, #2 ]}
\def\dotp[#1,#2]{(#1) \cdot (#2)}
\def\aff[#1,#2]{\hat{#1}(#2)}
\def\n4sym{{\cal N}=4 SYM}
\def\>{\rangle}
\def\<{\langle}
\def\weight[#1,#2,#3]{\{(#1),#2,#3\}}
\def\ads[#1]{$\text{AdS}_{#1}$}

  \makeatletter
  \let\over=\@@over \let\overwithdelims=\@@overwithdelims
  \let\atop=\@@atop \let\atopwithdelims=\@@atopwithdelims
  \let\above=\@@above \let\abovewithdelims=\@@abovewithdelims

\usepackage{xcolor}

\begin{document}

\preprint{}

\title{Pulsating flow and boundary layers in viscous electronic hydrodynamics}

\author{Roderich Moessner} 
\author{Piotr Sur\'owka}
\author{Piotr Witkowski} \affiliation{Max-Planck-Institut  f\"ur Physik komplexer Systeme, N\"othnitzer Str. 38, 01187 Dresden, Germany}

\begin{abstract}
Motivated by experiments on a hydrodynamic regime in electron
transport, we study the effect of an oscillating electric field in
such a setting. We consider a long two-dimensional channel of width
$L$, whose geometrical simplicity allows an analytical study
as well as hopefully permitting experimental realisation. 
The response depends on viscosity $\nu$, driving frequency,
$\omega$ and ohmic heating coefficient $\gamma$ via the dimensionless
complex variable $\frac{L^2}{\nu}(i\omega +\gamma)=i\Omega +\Sigma$. While
at small $\Omega$, we recover the static solution, a new regime
appears at large $\Omega$ with the emergence of a boundary layer.  This includes a
splitting of the location of maximal flow velocity from the centre towards
the edges of the boundary layer, an an increasingly reactive nature of the response, with the phase shift of the response varying across the channel. The scaling of the total optical
conductance with $L$ differs between the two regimes, while its
frequency dependence resembles a Drude form throughout, even in the
complete absence of ohmic heating, against which, at the same time,
our results are stable. Current estimates for transport coefficients in graphene and delafossites suggest that the boundary layer regime should be experimentally accessible.
\end{abstract}

\maketitle
\emph{Introduction.}---Hydrodynamics is an universal theory, that constantly reaches new areas of applicability. Recent experimental developments \cite{Molenkamp1994,deJong1995,Crossno2016,Bandurin2016,Moll2016,KrishnaKumar2017,Mackenzie2017} suggest that in clean materials electrons can acquire collective, hydrodynamic behavior. This is perhaps not so surprising since we expect that clean physical systems at large enough scales should behave as a fluid. In fact, first attempts to describe fluid behavior in electronic systems date back to Ghurzi in the 1960s \cite{Gurzhi1963,Gurzhi1968}. Experimental realizations of Ghurzi's ideas were, however, quite challenging because high impurity-related phenomena in most materials presented a formidable obstacle, with scattering of electrons off impurities and phonons spoiling the momentum conservation underpinning the hydrodynamic regime. Nevertheless, thanks to the advancements in chemical synthesis the number of candidate materials has grown to include two-dimensional (Al,Ga)As heterostructures, delafossite metals, and graphene \cite{Molenkamp1994,deJong1995,Crossno2016,Bandurin2016,Moll2016,KrishnaKumar2017,Mackenzie2017}. The relevant regime has sample size $L$ small enough that the mean-free path of electrons for the collisions with impurities and phonons $l_{MR}$ is larger than $L$. In graphene and delafossites, this separation of scales appears especially pronounced, thus making them ideal candidates to measure hydrodynamic signatures. In this case transport is dominated by momentum-conserving collisions. So far several examples of local \cite{Andreev2011,Alekseev2016,Scaffidi2017,Lucas2017} as well as non-local \cite{Torre2015,Levitov2016,Guo2017} responses of electronic flows have been identified. 

We ask the following basic question: what are the features of the electronic flow once we apply a periodic driving by an oscillating electric field? 
The simplicity of our set-up permits an analytic treatment of the solution and the concomitant experimental geometry is arguably the simplest possible. We identify a new high-frequency flow regime of the boundary-layer type. We interpret it as a genaralization of the so-called annular effect to charged electronic flows, although the name is more suited to three-dimensional flows through pipes \cite{Schlichting}. The annular effect was previously studied in fluid dynamics, in experiments with water driven by a periodically moving piston \cite{Richardson1929,Sexl1930,Uchida1956}. The emergence of a boundary layer provides a clear signature of a hydrodynamic behavior which results in several distinctive consequences. To facilitate the analysis we use dimensionless quantity $\Omega=L^2\omega/\nu$ (closely related to the Womersley number used in physics of pulsating flows) that depends on the sample size $L$, driving frequency $\omega$ and the kinematic viscosity $\nu$. This quantity can be thought of as a Reynolds number that controls various approximations in our set-up.

Our simple set-up is that of a long, sample in two-dimensions with non-slip boundary conditions. For a constant driving field, we can visualise it as a parallel flow (with only one velocity component) through the channel. In this case flow obeys an well-known solution, which corresponds to a 2 dimensional analogue of Hagen-Poiseuille flow in fluid mechanics. 

In the next sections we first introduce the theoretical set-up, with the Navier-Stokes equations in non-dimensional form subject to an oscillating electric field. We check that the known pulsating flow solution \cite{Langlois2014} is stable against addition of ohmic dissipation. We examine the behavior of the resulting flow profile as a function of $\Omega$ and dimensionless Ohmic coefficient $\Sigma$, identifying the boundary layer regime and analysing conditions under which it can be observed. We conclude that the boundary layer phenomena may be relevant for modern ''viscous electronic'' systems such as graphene or delafossites, and therefore are experimentally relevant. We calculate the conductance of the driven system and study its scaling behavior. We close the discussion with suggestions for experiments and an outlook.

\emph{Navier-Stokes equations.}---Fluid behavior can be reached for a given system if we probe the dynamics at scales that are large compared to the mean-free path of the microscopic constituents.  The equations of motion capture conservation laws of momentum, energy and charge 
\begin{equation}
\label{eq:cont}
\p _t \rho{} +\nabla _i (\rho u^i)=0,
\end{equation}
\begin{equation}
\label{eqn:system2d}
\p _t (\rho u^i)+\nabla_j (\rho u^i u^j)= \nabla_j \sigma ^{ij} +F^i.
\end{equation}
Here $u$ is the local fluid velocity, $\rho$ the fluid density, $F$ the force acting on the fluid, $\sigma ^{ij}$ the viscous stress tensor, and $\nu$ denotes kinematic viscosity. This is a very general set of equations so simplifying assumptions for the situation of interest are needed. We assume, similar to the static case, that the fluid is incompressible and isotropic. We also use the following ansatze:
\begin{equation}
u^y=u(t,x),~u^x=0,~F^y=\mu |E| e^{i\omega t},~F^x=0~~,
\end{equation}
where $y$ is the direction perpendicular and $x$ parallel to the channel, $\mu$ and $|E|$ stand for charge density and the intensity of the electric field respectively. The assumption on velocity as well as the fact that our flow should be viscosity-dominated allows us to neglect the non-linear advection term of \eqref{eqn:system2d}. Finally, we include a non-hydrodynamic, Ohmic dissipative term, with coefficient $\gamma$.  Then \eqref{eq:cont} and \eqref{eqn:system2d} reduce to a single equation
\begin{equation}
\label{eqn:OurEq}
\p_t u-\nu \p^2_x u +\gamma u = \mu |E| e^{i\omega t}.
\end{equation}
We non-dimensionalize our equation using 
\begin{equation}
u=L\omega{}u^\ast,~t\omega=\tau,~x/L=\chi~~,
\end{equation}
yielding dimensionless  velocity $u^\ast$,  relative width $\chi$, and time $\tau$ (measured in radians), 
Equation  \eqref{eqn:OurEq} becomes
\begin{equation}
\label{eq:dimlessEquation}
\p_\tau u^\ast - \frac{1}{\Omega} \p_\chi^2 u^\ast + \Gamma u^\ast = \Phi e^{i\tau}
\end{equation}
with
\begin{equation}
\Phi = \frac{\mu{}|E|}{\rho{}\omega^2 L},\qquad\Omega = \frac{L^2 \omega}{\nu},\qquad\Gamma = \gamma/ \omega.
\end{equation}
The first parameter encodes the forcing scale, the second is related to the Reynolds number in our electronic flow while the latter encodes the Ohmic friction coefficient.
Before we proceed to construct a solution we factor out the dependence on frequency in such a way that it is through the dimensionless ratio $\Omega$. This introduces new dimensionless parameters
\begin{equation}
\Psi = \Phi \Omega^2 = \frac{\mu E L^3}{\rho\nu^2},\qquad\Sigma=\Gamma\Omega=\frac{L^2\gamma}{\nu}.
\end{equation}
The velocity function depends on the spatial coordinate and time. Now the system can be solved analytically, and a general solution (for static initial condition $u(\tau=0)$) assumes form of an infinite series. However, the interesting 'late-time', periodic part can be extracted in a closed form by the use of the following substitution:
\be
u^\ast=\exp(i \tau) \mathcal{U}(\chi).
\ee
The above ansatz gives the following solution to Eq.\eqref{eq:dimlessEquation}:
\begin{equation}
\label{eq:blsol}
u^\ast=e^{i \tau}\frac{\Psi}{\Omega(i\Omega+\Sigma)}\left(1-\frac{\cosh(\sqrt{i\Omega+\Sigma}\chi)}{\cosh(\sqrt{i\Omega+\Sigma}/2)} \right).
\end{equation}
\begin{figure}[hbtp]
\centering
\includegraphics[width=0.47\textwidth]{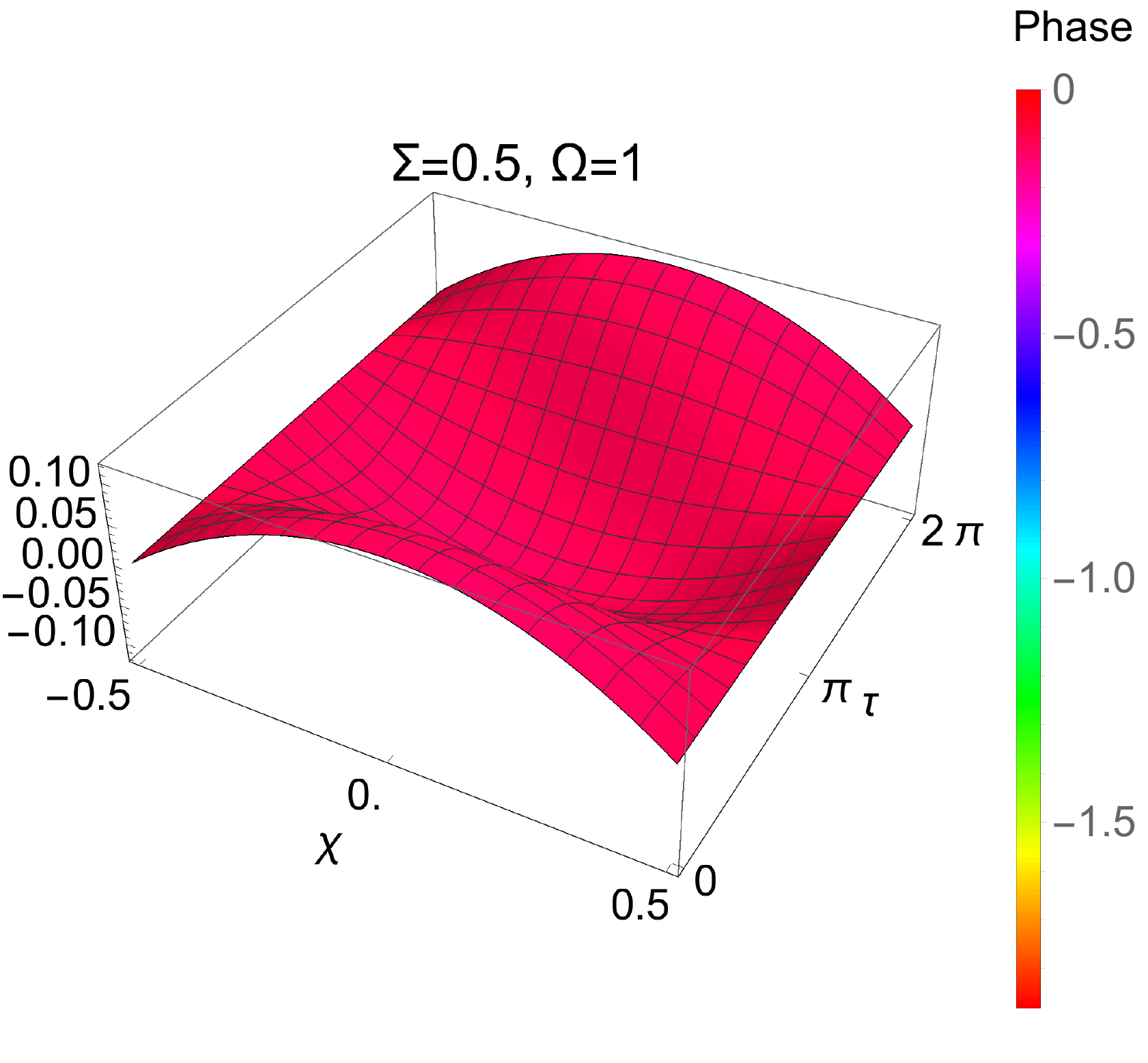} 
\includegraphics[width=0.47\textwidth]{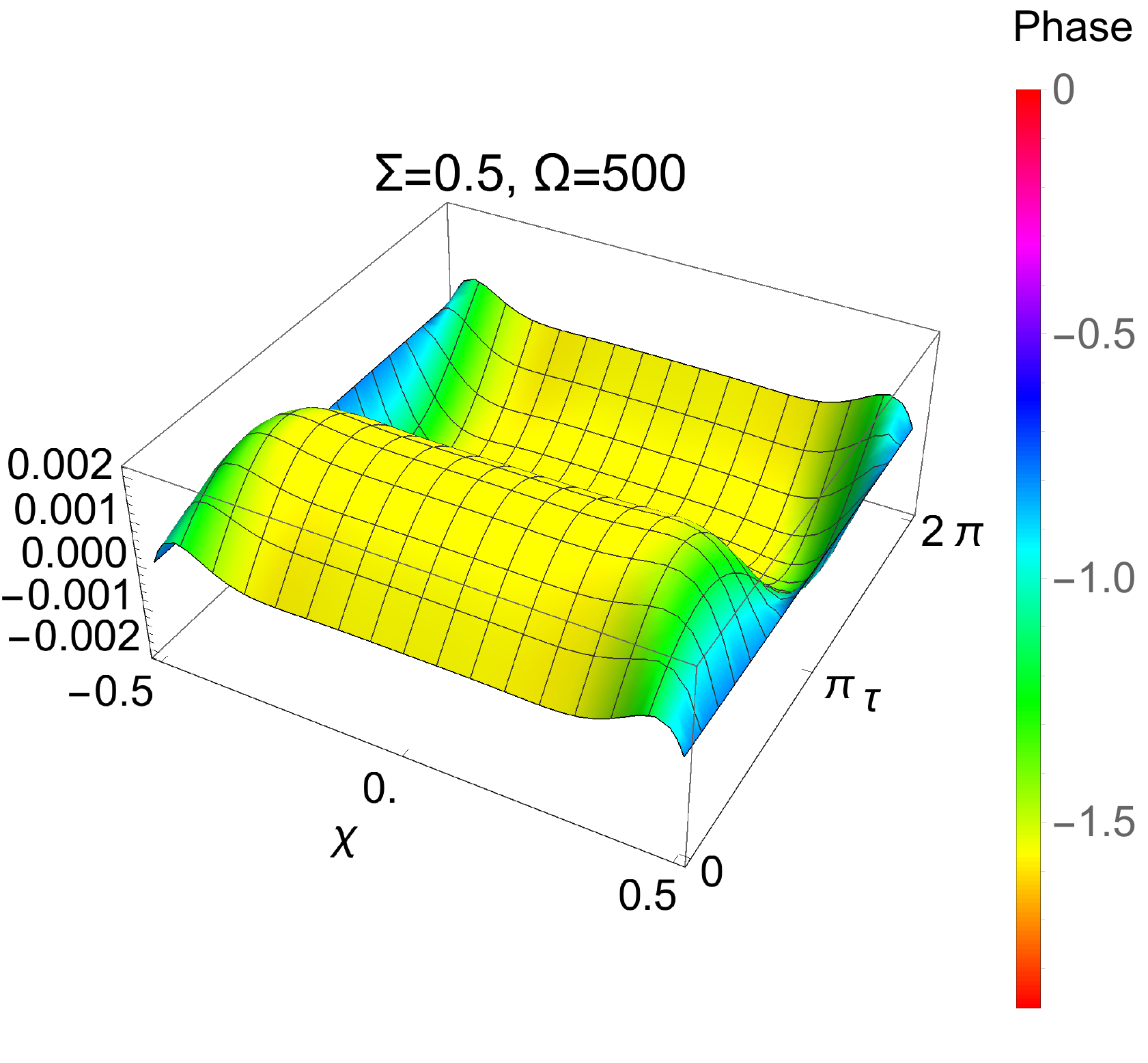} 
\caption{Spatiotemporal flow profiles $u^\ast (\chi, \tau)$ (z-axis) for different driving frequencies. The color represents the phase difference between the driving and the fluid velocity.}
\label{pic:Profile-SlowVFast}
\end{figure}
We can recover the physical velocity using the transformation $u=\frac{\nu}{L}\Omega u^\ast $. 

We first consider the asymptotic behavior for small and large frequencies. For the former, the velocity profile retains the same phase as the field that drives it, while the amplitude is parabolic as in the steady case. However, for large frequencies, the fluid ceases to be in phase with the rapidly oscillating force. In this case the solution acquires Stokes boundary-layer behavior. The fluid in the middle oscillates uniformly, and with an `inductive' phase relation to the drive, while close to the boundary it is dominated by  viscous effects. The velocity amplitude is plotted in Fig. \ref{pic:Profile-SlowVFast}. 
For slow forcing, the maximal velocity is reached in the middle of channel. Fast forcing induces the maximal velocity some distance from the channel center; the phase shift is visible, with the flow turning first near the boundaries and only then in the middle.

\emph{Conductance.}---Having the velocity profile $u$ we can write down the charge current $k=u\mu$
\begin{align}
\label{eq:define_j}
k&=\frac{\nu \mu}{L}\Re\left[ e^{i \tau}\frac{\Psi}{i\Omega+\Sigma}\left(1-\frac{\cosh(\sqrt{i\Omega+\Sigma}\chi)}{\cosh(\sqrt{i\Omega+\Sigma}/2)} \right) \right] \\ \nonumber
&\equiv \frac{\mu\nu}{L}\Psi \Re\left[e^{i \tau}j(\chi, \Omega, \Sigma) \right].
\end{align}
This is a local expression and we have to integrate over the sample width to extract conductance
\begin{equation}
\sigma = C\left[\frac{1}{\Sigma +i \Omega }-\frac{2 \tanh \left(\frac{1}{2} \sqrt{\Sigma +i \Omega }\right)}{(\Sigma +i \Omega )^{3/2}} \right],
\label{eq:OpticalConductivity}
\end{equation}
with $C=\frac{\mu^2 L^2}{\rho\nu}$, a dimensionful constant.
\begin{figure}[t]
\centering
 \includegraphics[width=0.45\textwidth]{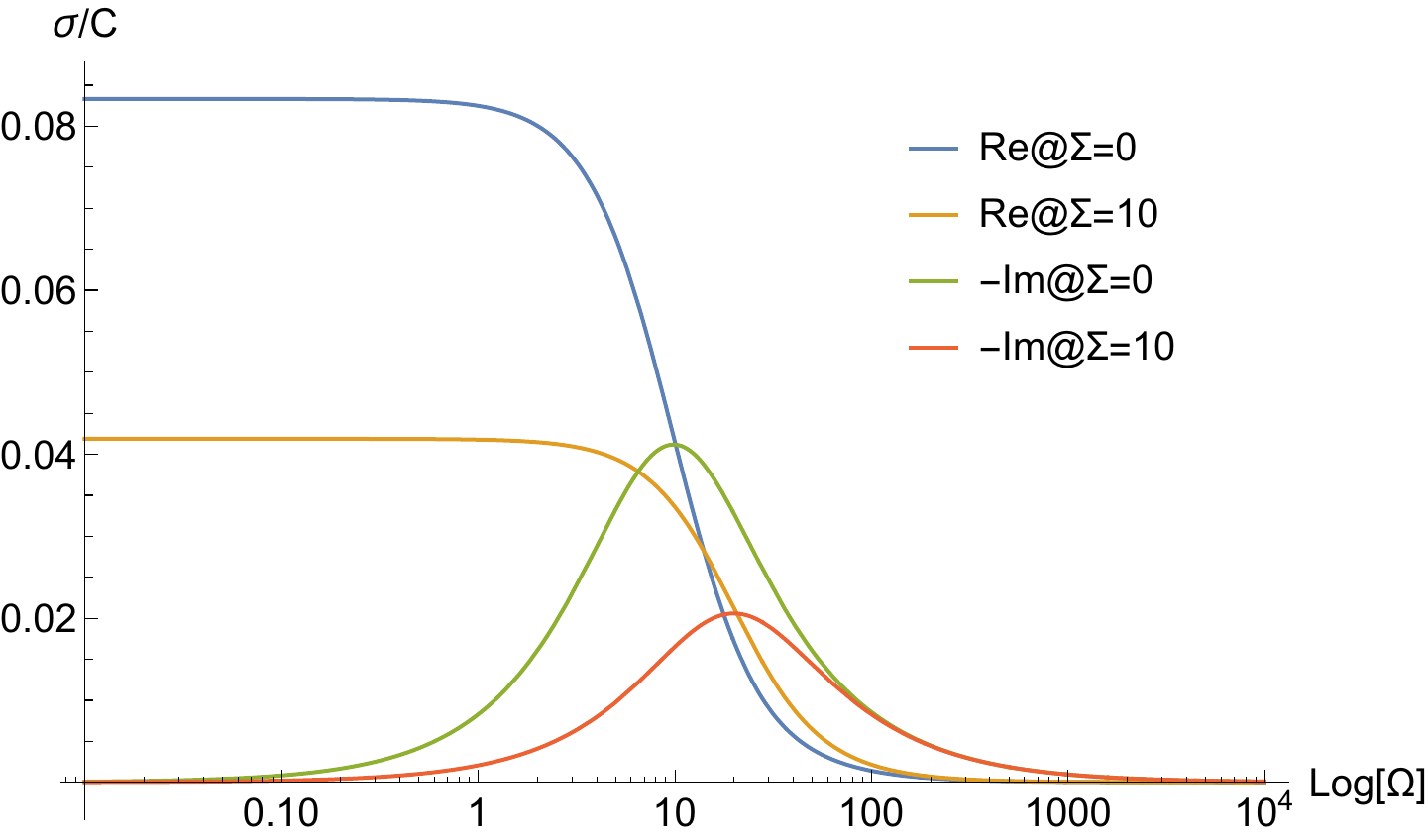} 
 \includegraphics[width=0.45\textwidth]{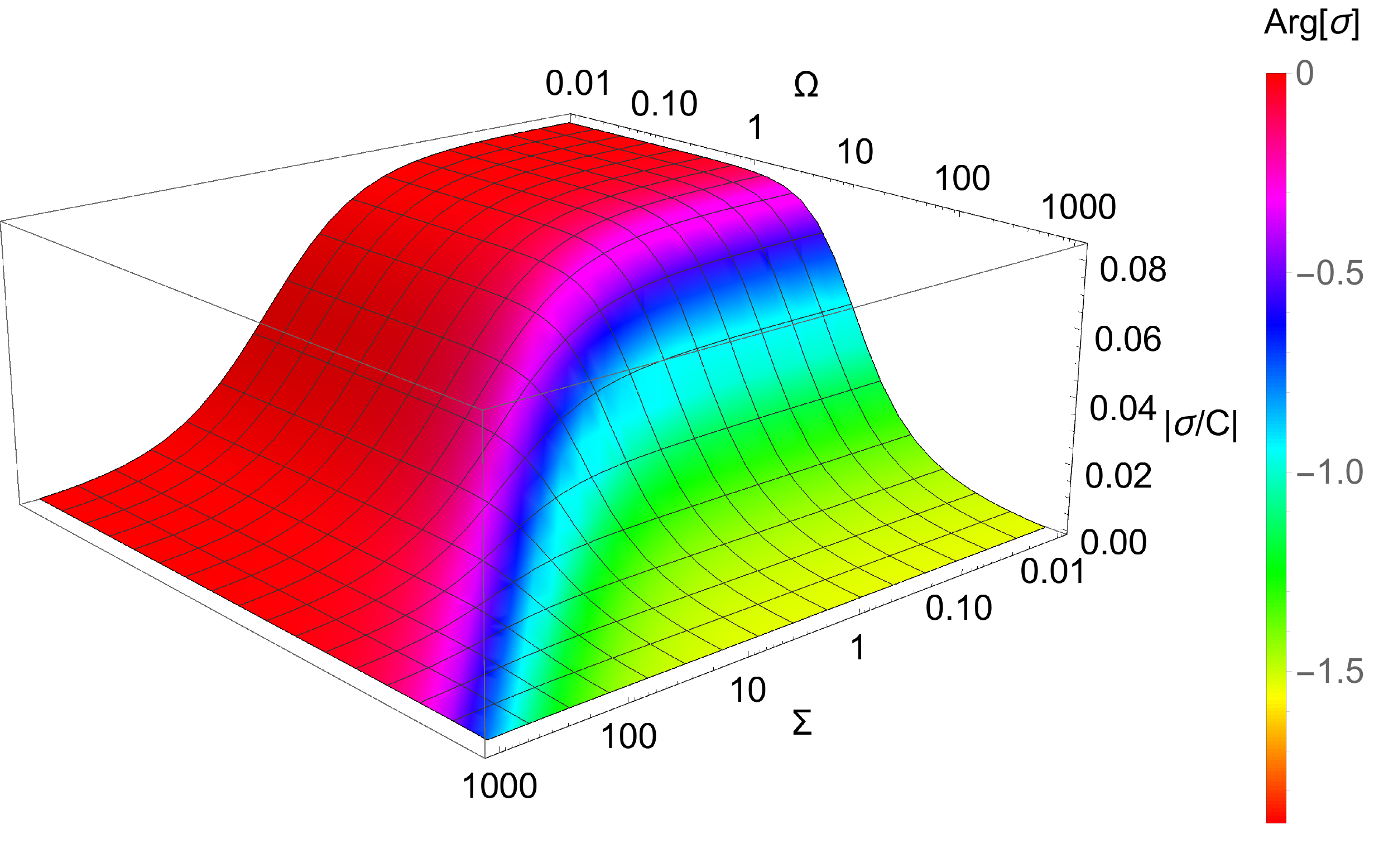}
 \includegraphics[width=0.45\textwidth]{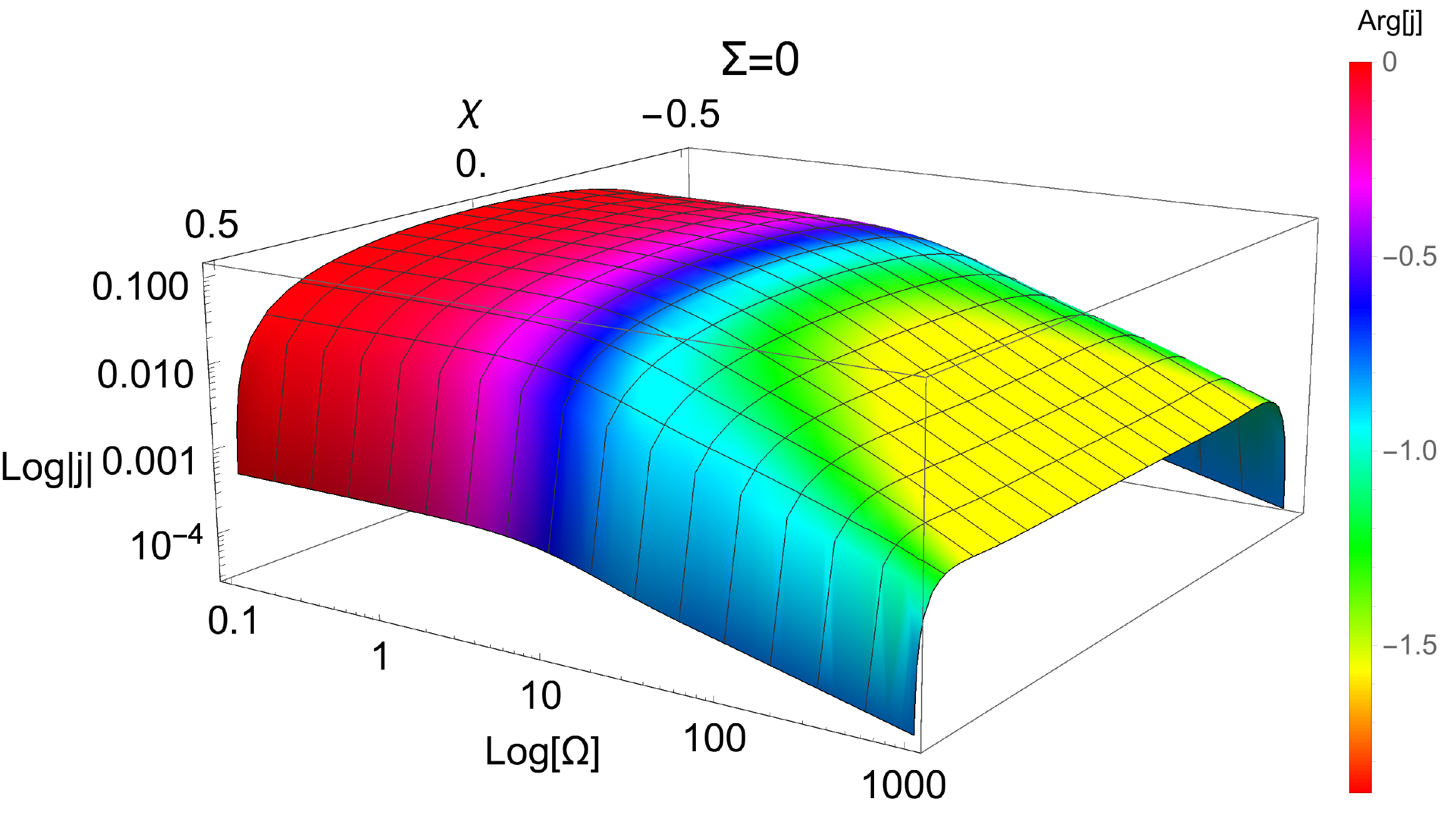}
\caption{Top: Real and imaginary parts of conductance for various choices of the parameter $\Sigma$. The bahavior resembles Drude conductivity and is stable against small Ohmic perturbations. The position of the Drude peak is $\Omega\approx{}9.89$ at $\Sigma=0$. The Drude peak separates two different scaling regimes, low frequency ($\sigma\sim\omega^0$) and high frequency ($\sigma\sim\omega^{-1}$).  Middle: Conductivity plotted in the $\Sigma-\Omega$ plane. The hydrodynamic behavior is stable for $\Sigma<10$. Bottom: local conductivity $j(\Sigma=0)$ as a function of the relative channel width $\chi$ and $\Omega$. Color encodes a local phase shift with respect to forcing, resistive at small $\Omega$, tuning reactive (inductive) at large $\Omega$.}
\label{pic:ReImCOnd}
\end{figure}
In order to identify the effect of driving on conductance, given the above solution, we have to possibilities: 1) fix the driving frequency $\omega$ and measure the conductivity for different sample sizes $L$, 2) vary the driving frequency at fixed sample size. It turns out that the effect of frequency changes is somewhat less striking in the flow profile. This follows from the scaling properties of the solution \eqref{eq:OpticalConductivity}, which is highly suppressed by $1/\omega$ upon re-scaling (see Fig. \ref{pic:ReImCOnd}). Thus we focus on the sample sizes. Due to the properties of the solution \eqref{eq:blsol} the re-scaling of length manifests itself only in the hyperbolic functions. As a result there is no suppression and we are able to extract a quantitative difference in conductance.

\emph{Scaling with the channel width.}---The global behavior of the current is modified if the constriction is small enough because the velocity profile differs significantly between non-viscous and boundary-layer regimes. As one can see in Fig. \ref{pic:Profile-SlowVFast} the driving creates a plateau between two ridges of the boundary layers. The thickness of the layer is fixed by the viscosity and the frequency. As a result, if the constriction is comparable to the width of the boundary layer the plateau dissapears and the flow qualitatively changes. Let's imagine we measure conductance for some width $L^\ast$ and fixed $\omega$. Then we measure for another sample whose width is $L=\beta L^\ast$. If we use asterisks to denote the reference values of our constants present in the formula for conductivity \eqref{eq:OpticalConductivity} then parameters scale with $\beta$ as:
\begin{equation}
\label{eq:WidthScaling}
\Omega = \beta^2 \Omega^\ast,~\Sigma=\beta^2\Sigma^\ast,~~C=\beta^2 C^\ast~~,
\end{equation}
whence the conductance
\begin{equation}
\sigma = C^\ast \left[\frac{1}{\Sigma^\ast +i \Omega^\ast }-\frac{2 \tanh \left(\frac{\beta}{2} \sqrt{\Sigma^\ast +i \Omega^\ast }\right)}{\beta{}(\Sigma^\ast +i \Omega^\ast )^{3/2}} \right].
\end{equation}
\begin{figure}[hbtp]
\centering
\includegraphics[width=0.4\textwidth]{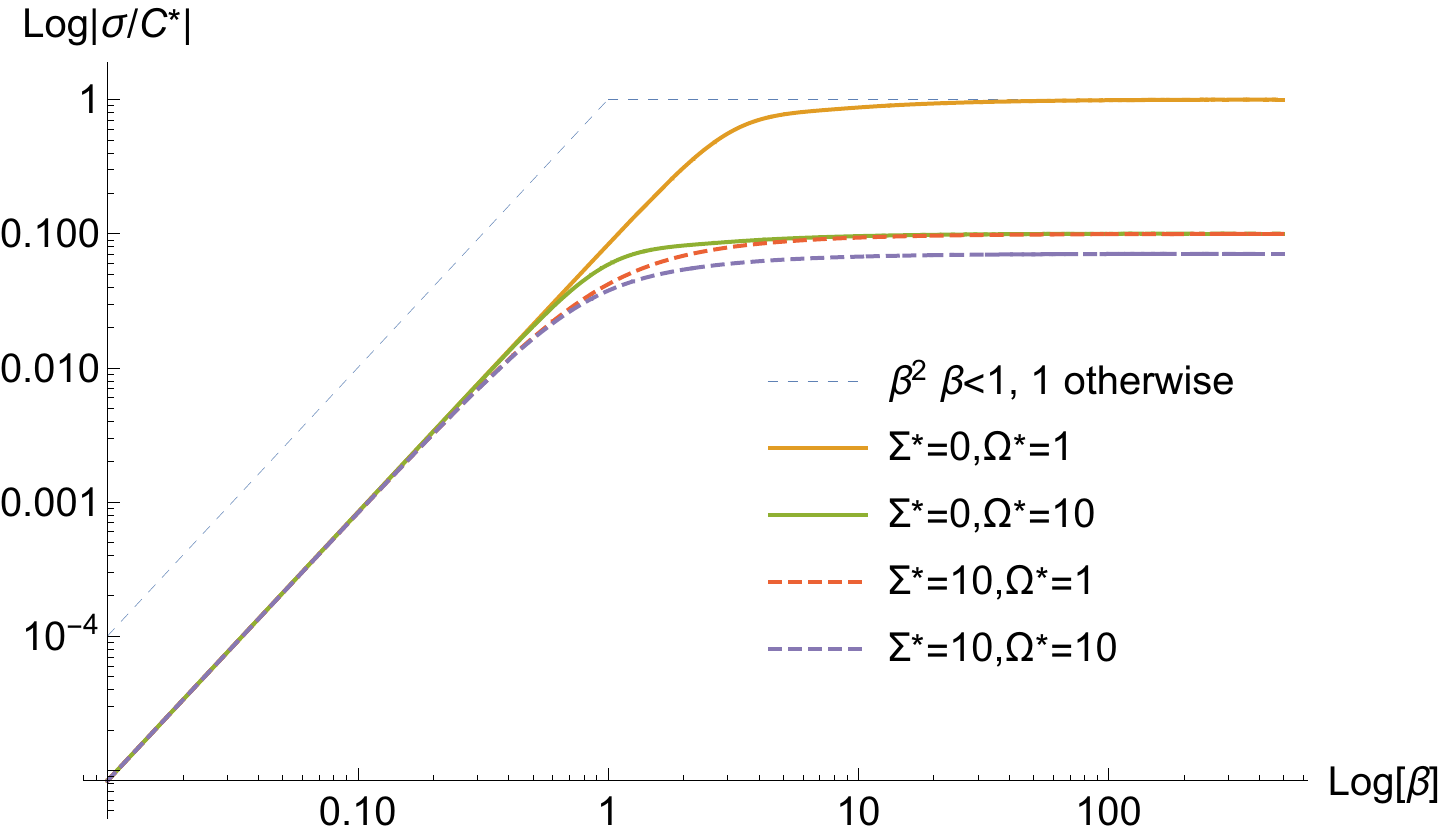} 
\includegraphics[width=0.4\textwidth]{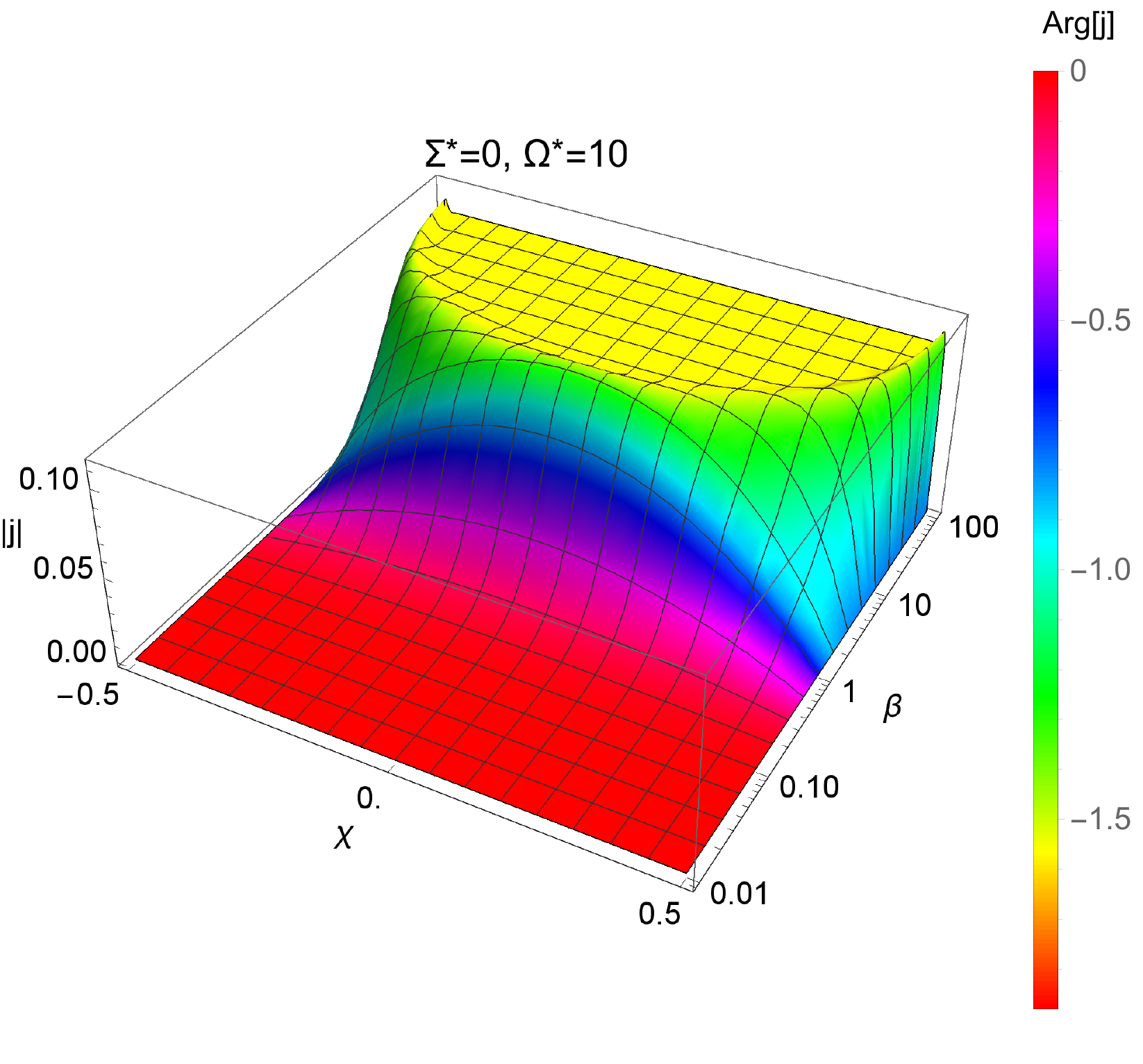}
\caption{Top: Scaling of the optical conductance for various values of the reference parameters $\Sigma^\ast$, $\Omega^\ast$. We set $C^\ast$ to 1. For $\beta$ smaller than the annular crossover value we observe $\sigma\sim\beta^2$ behavior. Bottom: local conductivity $j$ as a function of the width scaling $\beta$. The cross-over to annular flow happens approximately for the same value as the scaling change of the total conductance.}
\label{pic:WidthScaling}
\end{figure}This is plotted for several parameter values in Fig. \ref{pic:WidthScaling}. On a logarithmic scale we observe the initial linear growth of the conductance which then saturates. Note that Ohmic friction does not have a significant impact on the quantitative behavior.

\emph{Experimental implications.}---Pulsating flows have been experimentally investigated for three-dimensional ducts. In these experiments the local velocity measurements identified the boundary layers in the flow \cite{Richardson1929}, to which at high-frequency the effects of viscosity are confined. We now discuss under what conditions a boundary-layer emerges in electronic systems. Our analysis suggests a very simple experimental set-up, which consists of a longitudinal sample of e.g. monolayer graphene positioned between voltage probes. On top of that we require an external drive. Our considerations are based on experiments in graphene or delafossites, in which the current technology allows one to reach sample sizes around $0.01$ of the electron mean free path, for $L \approx 10^{-6} m$. In the case of delafosite metals this can be achieved from flux-grown single crystals
by focused ion beam etching \cite{Moll2016}. Bilayer graphene devices are prepared crystals of hexagonal boron nitride using lithography and subsequent etching processes \cite{Bandurin2016}. The estimates for kinematic viscosity are around $\nu =0.1 m^2/s$ \cite{Crossno2016,Bandurin2016,Moll2016,KrishnaKumar2017,Mackenzie2017}. The boundary layer requires high frequencies $\Omega \gg 1$. Given the above estimates for $\nu$ and $L$ we deduce the boundary-layer to have emerged at $\Omega = 100$, around 100 GHz, i.e. in the high-frequency microwave/far infrared part of the electromagnetic spectrum. Numerical evaluation of the solution \eqref{eq:blsol} suggests that the layer actually starts to emerge at  $\Omega \approx 80$. Of course one may worry if this lies in the range of applicability of hydrodynamic description. However, comparing the time scales: momentum-conserving collision rate $\gamma_{MC}\approx 8\times{}10^{-14}s$ \cite{Levitov2016} and forcing frequency time scale $100GHz$ ($\approx{}10^{-11}s$) leads us to the conclusion, that we are still deep in the regime, where momentum-conserving interactions are dominating enough to facilitate the use of collective (i.e hydrodynamic) description. 
Standard estimates for the width of the boundary layer give
\begin{equation}
\delta \sim \sqrt{\frac{\nu}{ \omega}},
\end{equation} 
which, for the values of viscosity measured for graphene or delafossite metals, results in the layer thickness of order $10^{-7}m$. The parameter $\Sigma$ produced by Ohmic friction can reach the numerical values up to 10, based on the experimental estimates \cite{Bandurin2016,Moll2016,Mackenzie2017}. 

As we already mentioned, the given analysis focuses on periodic solutions, i.e. ones with real frequency. One can also ask how is this periodic steady state reached. To answer that one must solve equation \eqref{eq:dimlessEquation} on $\chi\in{}[-1/2;1/2],~\tau\geq{}0$. The analytical solution consists of the periodic part \eqref{eq:blsol} and an infinite series of functions of complex frequency which correspond to exponentially dumped modes. The slowest-relaxing mode has a lifetime given by 
\begin{equation}
t_{rel} = \frac{\nu \pi^2}{L^2} \approx{} 10^{-12}s,
\end{equation}   
with the last number estimated using previously mentioned parameters. The non-periodic behavior is therefore experimentally irrelevant.

\emph{Discussions.}---We have analysed the behavior of the electronic fluid driven by a
periodic voltage difference. If the driving frequency is large enough, the system presents a clear
hydrodynamic signatures in the form of a boundary-layer structure of the
flow. The boundary layer emerges as a consequence of viscosity and the
boundary conditions, thus giving a distinctive hydrodynamical feature absent
in both Ohmic and ballistic regimes. 

Studying the purely hydrodynamic regime we have identified several consequences
of the boundary-layer flow, the most prominent being the
dependence of the conductivity on  sample size and frequency. The simplicity of our
considerations allows for clear experimental protocols, which should be realistically
possible with current experimental capabilities, to seek
further evidence for the hydrodynamic behavior of electrons in solids.
AC susceptibility measurements are staple electronic experiments and the scaling properties (Fig. \ref{pic:WidthScaling} and \ref{pic:ReImCOnd}) discussed in the previous section should provide a 
global signature of hydrodynamic flow.
More ambitiously, it would be gratifying to image the boundary layer directly using local probes, although it is not clear whether direct measurements of such probes (ideally -- current densities) are available at this point of time.

The existence of a boundary layer also  has "technical" implications, e.g. for numerical treatment of this and similar systems in more complex approaches.
With a thin layer crucially impacting the global response, it will be necessary to choose a set-up -- e.g. with a spatially dependent grid size -- capable of capturing this fine structure in simulations,
which could be applied in the framework of kinetic theory in order to work towards a complete understanding of the ballistic-hydrodynamic crossover. This would require solving a Floquet-Boltzmann equation \cite{Genske2015} with
the Bhatnagar-Gross-Krook collision term \cite{Bhatnagar1954}.

In addition, we note that studying boundary layers in fluid flows is an
important subject in its own right.
Indeed, the appearance of boundary layer physics is presumably not limited to the AC-drive in a channel, described here, and it will be interesting to look for similar phenomena in other appropriate settings.
From that perspective electronic systems are unique test objects, as (contrary to the classical, mechanically driven fluids) they are forced by the electric field, which is highly tunable, can easily be made uniform and propagates with the speed of light (effectively instantaneously in the hydrodynamic approximation). 
Therefore, two-dimensional boundary-layer electronic flows can be used as a new test bed of for older ideas, and allow the study of vorticity and the emergence of turbulent behavior.

\emph{Acknowledgments}---We acknowledge useful conversations with Renato Dantas, Andy Mackenzie, Takashi Oka, Kush Saha, Thomas Scaffidi, and J\"{o}rg Schmalian. This work was supported by the Deutsche Forschungsgemeinschaft via the Leibniz Programm.

\bibliographystyle{apsrev4-1}

\bibliography{floquet-bib}

\end{document}